\documentclass[english,twocolumn,aps,pra,longbibliography,crop,tikz,superscriptaddress]{revtex4-1}
\usepackage[T1]{fontenc}
\usepackage{graphicx}
\usepackage{babel}
\usepackage{braket}
\usepackage{qcircuit}
\usepackage{xcolor}
\usepackage{amsmath}
\usepackage{lipsum}
\usepackage{blindtext}
\usepackage{upgreek}
\usepackage{soul}
\usepackage{multirow}
\usepackage{mathtools}
\usepackage{array}

\usepackage{qcircuit}

\usepackage[colorlinks,linkcolor=black,citecolor=black,urlcolor=black,bookmarks=false,hypertexnames=true]{hyperref} 

\usepackage{fancyhdr}

\fancypagestyle{fancyplain}
\newcolumntype{R}[1]{>{\raggedleft\arraybackslash}p{#1}}

\begin{document}
\pagestyle{fancy}

\title{Method for simulating open-system dynamics using mid-circuit measurements on a quantum computer}

\author{John P. T. Stenger}
\thanks{Corresponding Author: john.stenger@nrl.navy.mil}
\affiliation{Chemistry Division, U.S. Naval Research Laboratory, Washington, DC 20375, United States}
\author{Gloria Bazargan}
\affiliation{Chemistry Division, U.S. Naval Research Laboratory, Washington, DC 20375, United States}
\author{Nicholas T. Bronn}
\affiliation{IBM Quantum, IBM T.J. Watson Research Center, Yorktown Heights, New York 10598, USA}
\author{Daniel Gunlycke}
\affiliation{Chemistry Division, U.S. Naval Research Laboratory, Washington, DC 20375, United States}

\begin{abstract}
We present a method for simulating the dynamics of an open electronic system on a quantum computer. This approach entails mid-circuit measurements and resets to simulate the addition or removal of electrons from the system.  Our method provides a way to apply non-reversible operations to a quantum computer without the need for additional qubits. Using this method, we simulate the dynamics of an open electronic system consisting of a chain of electrons positioned between two conductive leads on the $ibm\_torino$ quantum computer.  We expect the method to be generally applicable to open systems.  
\end{abstract}

\maketitle
\thispagestyle{fancyplain}

\section{Introduction}

Quantum computing shows promise for simulating electronic systems.  This is due to the straightforward mapping between the unitary operators that govern electron dynamics and those describing the quantum logic gates that are executed on quantum computers. Consequently, there are many studies that report simulations of closed-system electron dynamics on quantum computers~\cite{Peruzzo2014,McClean2016,OMalley2016,Kandala2019,Wang2019,Brian2019,Arute2020,McArdle2020,HeadMarsden2021,Kim2023, Wang2024}.  Conversely, it is non-trivial to simulate the dynamics of open quantum systems on a quantum computer.  This is largely due to the fact that an open quantum system can have non-reversible dynamics while quantum logic gates are necessarily reversible.  Therefore, there are fewer reports of open-system electron dynamics simulations on quantum computers~\cite{Wei2018,Sweke2015,Candia2015,Wei2016,Wang2013,Wang2011,Hu2020,Bian2023,Li2023}. 

Simulating open systems is essential to understanding fundamental processes in chemical and biological environments, materials, and nanodevices. Examples of these processes include non-equilibrium phase transitions~\cite{Prosen2011}, electron and ion transport dynamics in proteins~\cite{Huelga2013,Mostame2012}, the preparation of thermal states~\cite{Terhal2000,Stenger2024}, light harvesting in nanodevices~\cite{Scully2011,Creatore2013}, tunneling spectroscopy~\cite{Stenger2022}, quantum to classical transitions~\cite{Zurek2003}, and determining the Markovian character of quantum transport~\cite{Breuer2009,Rivas2010}.

One strategy to simulate open quantum systems is to divide the quantum register into two parts, representing the system and the bath~\cite{Poulin2009,Wang2011,Anschuetz2019,Wu2019,Yuan2019,Su2020,Wang2021,Zoufal2021}. The non-reversible operations can then be applied to the system by tracing out the bath degrees of freedom using quantum measurement outcomes.  Other strategies include collecting ensembles of pure states~\cite{Cohn2020,holmes2023}, tracking the norm on a classical computer~\cite{Chen2024}, using knowledge of the correlation length~\cite{Motta2020}, and decomposing operators~\cite{Schlimgen2021}.

In this article, we present a method to simulate non-reversible operations on an open quantum system using mid-circuit measurements~\cite{Stenger2024} and resets . This method does not require any additional qubits to represent the bath degrees of freedom.  We apply this method to an open system modeling a conductive chain positioned between two electronic contacts: a source and a drain.  Our results show its simulated electron dynamics on a quantum computer and indicate the applicability of this method for studying the transport dynamics of open systems in general. 

\section{Methods}

\subsection{Closed electronic systems} 

We consider an electronic system described by a time-independent Hamiltonian $\hat H$ acting on a Fock space.  The evolution in time $t$ for a quantum state of an isolated system is described by the unitary operator
\begin{equation}
    \hat U(t) = e^{-i\hat H t}.
\end{equation}

\noindent This time evolution can be simulated on a quantum computer by mapping the Fock space to the Hilbert space for the quantum computer, using for example the Jordan--Wigner transformation~\cite{Jordan1928}. The unitary operator can be decomposed as a product of exponentials of Pauli operators using the Lie--Trotter approximation 
\begin{equation}
    \hat U(t) \approx  (\Delta \hat{U})^{N_t},
    \label{eq2}
\end{equation}
\noindent where $\Delta \hat U = \prod_s e^{-i\hat P_s t/N_t}$, $N_t$ is the number of time-steps and $\hat P_s$ are Pauli strings corresponding to terms in the Hamiltonian.  The Lie--Trotter approximation is needed because the $\hat P_s$ do not commute in general.  The operator $\Delta \hat U$ can be implemented using quantum logic gates~\cite{mansky2023}. This procedure is well-established and has been applied to electronic simulations on quantum computers~\cite{Abrams1997,Whitfield2011,Stenger2023} . Simulating the dynamics of open systems, however, is less straightforward and demands a strategy to implement non-reversible operations. 

\subsection{Open systems} 
\begin{figure}[h]
\vspace{2mm}
\begin{center}
\includegraphics[width=\columnwidth]{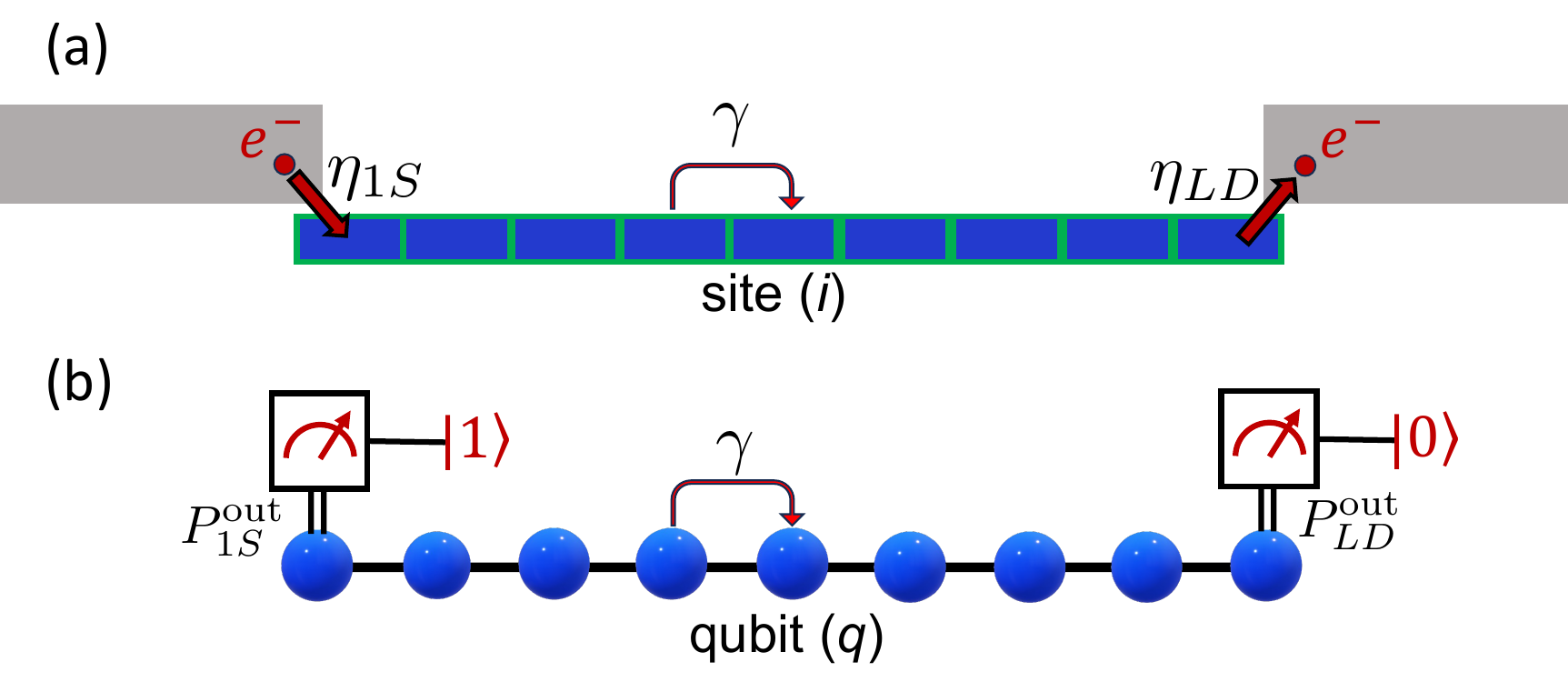}
\end{center}
\vspace{-2mm}
\caption{Schematic representation of the open electronic system and its representation on a quantum computer:  a) a nanowire (blue bar) in contact with an electronic source (left gray bar) and an electronic drain (right gray bar), b) the system is simulated with a defined number of qubits (blue spheres) where the electronic source (drain) are simulated by resetting the left-most (right-most) qubit to $\ket{1}$ ($\ket{0}$).}
\label{F0}
\vspace{-3mm}
\end{figure}

Consider an open system of electrons coupled to metallic contacts.  An example of such a system is a chain of electrons between source and drain contacts as illustrated in Fig.~\ref{F0}a.  We represent each site by a qubit, where the end qubits are connected to the contacts as shown in Fig.~\ref{F0}b.  Transport is modeled by evolution of the system state, in which electrons are injected or removed at the contacts.   To simulate the dynamics we introduce the state $\ket{\Phi(t)}$ at time $t$, starting from the initial state $\ket{\Phi(0)} = \ket{\Phi_0}$.  

 At each end site and time-point, we probabilistically select between three possible actions: (1) inject an electron, (2) remove an electron, (3) do nothing.  Consider a quantum register with $L$ qubits and let $\ket{0}_q$ and $\ket{1}_q$ be the basis states for  Hilbert space of qubit $q$.  The actions are performed through mid-circuit measurements.  An electron is injected by setting the qubit measured in $\ket{0}_q$ to $\ket{1}_q$ and removed by setting the qubit measured in $\ket{1}_q$ to $\ket{0}_q$.  For each end qubit $q$ and conductor $\alpha$, we define 
\begin{equation}
    \begin{split}
        &P^{\text{in}}_{q\alpha} = \eta_{q\alpha}f(\mu_{\alpha})
        \\
        &P^{\text{out}}_{q\alpha} = \eta_{q\alpha}[1 - f(\mu_{\alpha})],
        \\
         &P_{q\alpha} = 1 - P^{\text{in}}_{{q\alpha}} - P^{\text{out}}_{{q\alpha}} 
    \end{split}
\end{equation}
where $P^{\text{in}}_{q\alpha}$ and $P^{\text{out}}_{q\alpha}$ are the probabilities of electron injection and removal, respectively, $f$ is the Fermi--Dirac distribution function, $\mu_{\alpha}$ is the electrochemical potential of contact $\alpha$,  and $\eta_{q\alpha} = \Gamma_{q\alpha} t/N_t $, where $\Gamma_{q\alpha}$ is the tunneling rate~\cite{Maslova2021,Liu2025} between qubit $q$ and conductor $\alpha$.  Note that $N_t$ must be sufficiently large such that the probability of electron injection or removal is less than or equal to one, $N_t \geq (\Gamma_{q\alpha} t)^{-1}$ for all $\Gamma_{q\alpha}$.  In the limit $N_t \rightarrow \infty$, this method simulates the dynamics of a system in contact with two conductors.  See Appendix~\ref{rle} for a comparison of our method with the Lindblad equation \cite{Brasil2013}.

\subsection{Algorithm} 
Our algorithm is summarized in Table~\ref{T1}.  We have a list of qubits $q \in \{1,L\}$ and conductors $\alpha \in \Omega$.
We initialize the quantum register to the state $\ket{\Phi_0}$ such that the qubit for site $i$ is $\ket{1}$ if it contains an electron and $\ket{0}$ if it does not.  We apply the unitary operator $\Delta \hat U$ iteratively $N_t$ times as shown in Eq.~\eqref{eq2}.  After each application of  $\Delta \hat U$, we generate a random number $P$ between 0 and 1 use it to check the probability conditions. With probability $P < P^{\text{in}}_{q\alpha}$ we attempt to inject an electron into the system at qubit $q$ from conductor $\alpha$.  If $P^{\text{in}}_{q\alpha}<  P < P^{\text{in}}_{q\alpha} + P^{\text{out}}_{q\alpha}$ we attempt to remove an electron from the system from qubit $q$ into conductor $\alpha$.  With probability $P_{q\alpha} = 1- \eta_{q\alpha}$ nothing is done. The function $M(q,\ket{\Phi})$ represents a measurement of $\ket{\Phi}$ for qubit $q$.  The combined process of measuring qubit $q$ followed by conditionally switching the state to $\ket{0}_q$ is equivalent to reseting the qubit.  Similarly, measuring qubit $q$ and then conditionally switching the state to $\ket{1}_q$ is equivalent to reseting the qubit and then applying $\hat X_q$, the Pauli-X gate .         

\begin{table}[h]
\caption{\label{tab:example1} Pseudocode}
\begin{ruledtabular}
\begin{tabular}{ll}
1 : & $\ket{\Phi} = \ket{\Phi_0}$
\\
2 : & repeat $N_t$ times :  
\\
3 : & \quad \quad $\ket{\Phi} =\Delta \hat U \ket{\Phi}$  
\\
4 : & \quad \quad for $q \leq L$:
\\
5 : & \quad \quad \quad \quad for $\alpha \in \Omega$:
\\
6 : & \quad \quad \quad \quad \quad \quad $ P = \text{random(0,1)}$
\\
7 : & \quad \quad \quad \quad \quad \quad    if $P < P^{\text{in}}_{q\alpha}$:
\\
8 :  &\quad \quad  \multirow{3}{*}{reset($q$)  $\begin{dcases*} \\ \\ \end{dcases*}$} \quad     state $=$ $M(q,\ket{\Phi})$
\\
9 : & \quad \quad \quad \quad \quad \quad \quad \quad if state $== \ket{0}_q$:
\\
10 : & \quad \quad \quad \quad \quad \quad \quad \quad \quad \quad $\ket{\Phi} = \ket{1}\!\bra{0}_q \ket{\Phi}$
\\
11 : & \quad \quad \quad \quad \quad \quad else if $P^{\text{in}}_{q\alpha} < P < P^{\text{in}}_{q\alpha} + P^{\text{out}}_{q\alpha}$:
\\
12 : & ~~\multirow{3}{*}{$\hat X_q$ reset($q$)  $\begin{dcases*} \\ \\ \end{dcases*}$} \quad state $=$ $M(q,\ket{\Phi})$
\\
13 : & ~\quad \quad \quad \quad \quad \quad \quad \quad if state $== \ket{1}_q$:
\\
14 : &\quad \quad \quad \quad \quad \quad \quad \quad \quad \quad $\ket{\Phi} = \ket{0}\!\bra{1}_q \ket{\Phi}$   
\end{tabular}
\end{ruledtabular}
\label{T1}
\end{table}

\section{Demonstration} 

\subsection{Model System} 

We now demonstrate our method for the chain of interacting electrons, described by the Hamiltonian,
\begin{equation}
    \hat{H} = \gamma\sum_{<i,j>} \sum_\sigma (\hat c^{\dagger}_{i\sigma} \hat c_{j\sigma} + \hat c^{\dagger}_{j\sigma} \hat c_{i\sigma}) + v\sum_{<i,j>}  \sum_\sigma \hat n_{i\sigma} \hat n_{j\sigma},
\end{equation}
where $\hat c^{\dagger}_{i\sigma}$ creates an electron on site $i \in \{1,...,L\}$ with spin $\sigma \in \{\uparrow,\downarrow\}$, $\hat n_{i\sigma} = \hat c^{\dagger}_{i\sigma} \hat c_{i\sigma}$ is the density operator, 
$\gamma$ is the hopping integral, $v$ is the electron-electron interaction strength, and $<i,j>$ indicates that $i$ and $j$ are nearest neighbors.  We have chosen a model in which the two spin sectors are identical and decoupled.  Therefore, we can safely choose a spin sector and drop the spin index.  The electron movement across the chain is tracked by the time-dependent electron density $n_i(t) = \braket{\Phi(t)|\hat n{_i}|\Phi(t)}$. 
Using the Jordan-Wigner transformation~\cite{Jordan1928,Ortiz2001}, we find the transformed Hamiltonian

\begin{equation}
  \begin{split}
     \hat H =  \frac{\gamma}{2} \sum_{<q,p>} (\hat X_q \hat X_p + \hat Y_q \hat Y_p) + \frac{v}{4} \sum_{<q,p>} (\hat I - \hat Z_q)(\hat I - \hat Z_p).\\
  \end{split}
\end{equation}
where $\hat X_q$, $\hat Y_q$, and $\hat Z_q$ are Pauli operators acting on $q$.


We then simulate the electron chain both as a closed system, and as an open system coupled to contacts.  In the closed-system case, we set $\eta_{q\alpha} = 0$ for all $q$ and $\alpha \in \{\text{S},\text{D} \}$.  In the open-system case, we couple the system to a source $f(\mu_\text{S}) = 1$ at the first site $\eta_{1\text{S}} \neq 0$ and a drain $f(\mu_\text{D}) = 0$ at last site $\eta_{L\text{D}} \neq 0$, where $i=L$ is the last site.   

\subsection{Closed-system dynamics}
\begin{figure}[h!]
\vspace{2mm}
\begin{center}
\includegraphics[width=\columnwidth]{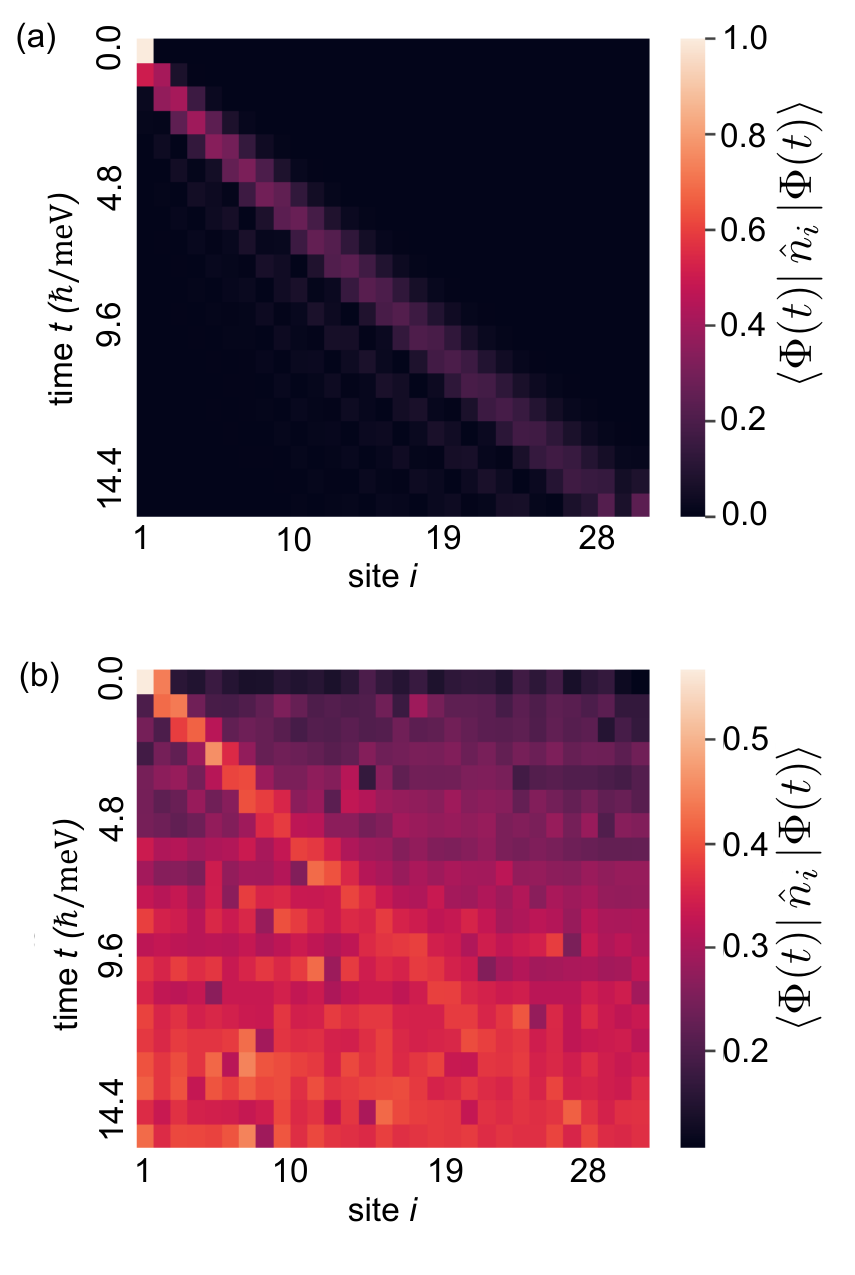}
\end{center}
\vspace{-2mm}
\caption{ A single electron traveling from site $i=1$ to site $i=30$.  (a) classical computer simulation. (b) quantum computer simulation. The time-dependent electron densities are shown, where light regions correspond to high density and dark regions correspond to low density. }  
\label{F1}
\vspace{-3mm}
\end{figure}

We begin by comparing the dynamics of a closed electronic system simulated on a classical computer to those simulated on a quantum computer. Figure ~\ref{F1} shows a model system containing a single electron.  We set  $L=30$, $\gamma = 1$ meV, $v = 0$, and $N_t/t = 2 ~\text{meV}/\hbar$.  For the simulation on the classical computer shown in Fig. 2a, the initial electron density at the left end is $n_1(0)=1$. The time evolution of this localized electron is described by Eq.~\eqref{eq2}.  We calculate $n_i(t)$ for each site and discrete time-point $t$. As shown in Fig.~\ref{F1}a, $n_1(t)$ decreases, while the density at neighboring sites increases. At $t=15 \hbar ~\text{meV}^{-1}$, there is accumulation of density at end site $i=L$. 

Now let us consider the dynamics of the same single-electron system on a quantum computer.  We run the simulation on the $ibm\_torino$ quantum computer.  See Appendix~\ref{apexb} for specifications of the quantum computer.  The results are shown in Fig.~\ref{F1}b. Similar to the simulation performed on the classical computer, the dynamics on the quantum computer show an increase in density at sites near the right end of the chain over time.  The noise in the electron density distribution in Fig.~\ref{F1}b compared to Fig. ~\ref{F1}a, is a consequence of quantum errors, including decoherence. 

\subsection{Open-system dynamics}
\begin{figure}[h!]
\vspace{2mm}
\begin{center}
\includegraphics[width=\columnwidth]{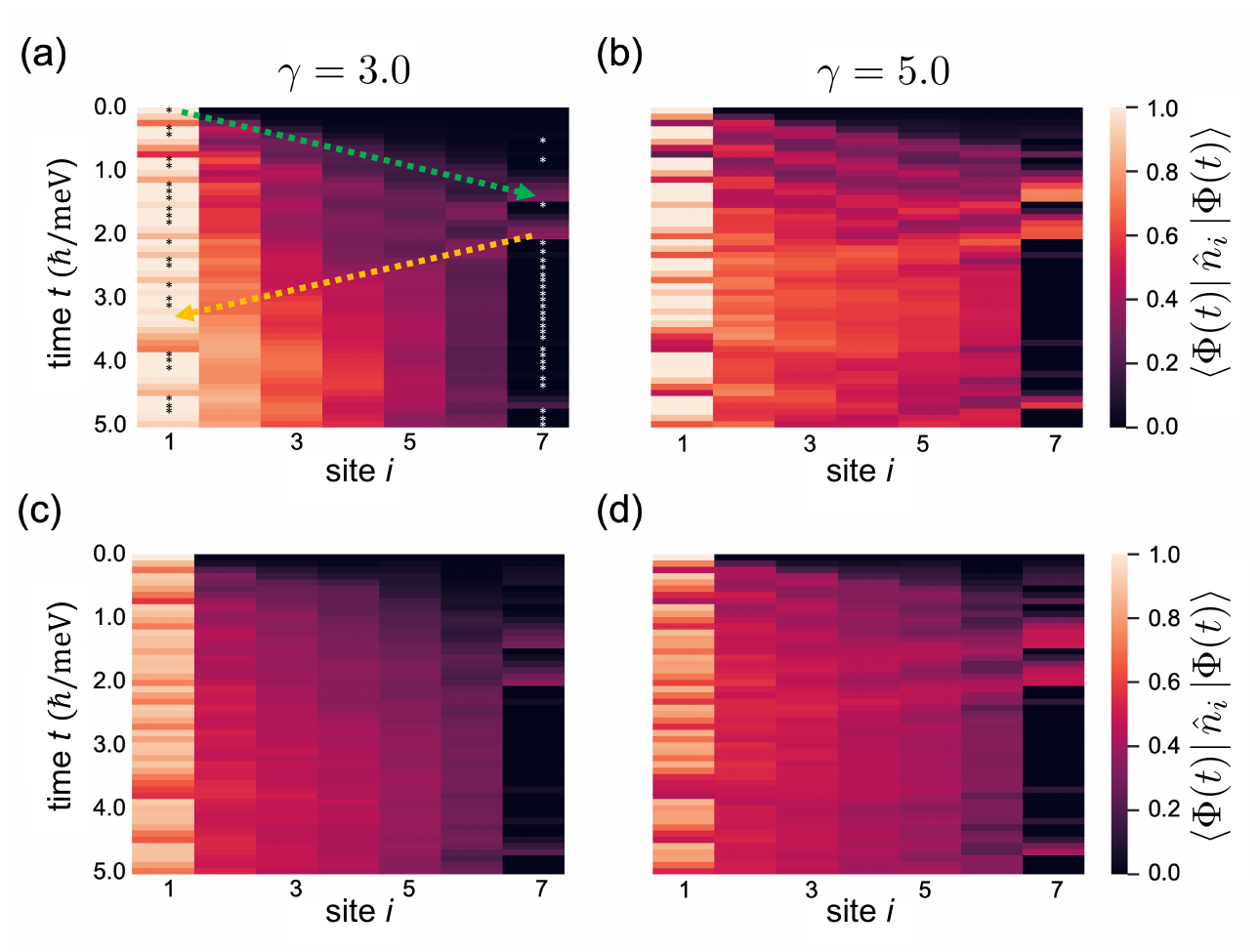}
\end{center}
\vspace{-2mm}
\caption{Simulations of open-system dynamics with $\gamma$=3.0 meV, and 5.0 meV: (a-b) classical computer, (c-d) quantum computer. The source and drain are at $i=1$ and $i=7$, respectively.  In (a) the points in which electrons are injected (removed) are indicated by black (white) asterisks.  The leading edge of electron density is marked by the dotted green arrow.  Peaks in the electron density corresponding to an electron reflected off of the right end is marked by the dotted orange arrow.  }  
\label{F3}
\vspace{-3mm}
\end{figure}

We now compare the dynamics of the open system simulated on a classical computer to those simulated on a quantum computer.  Figure~\ref{F3} shows the dynamics of a seven-site system as simulated on a classical (a-b) and quantum(c-d) computer.  We consider $\gamma$=3.0 meV and 5.0 meV, and set $v$=10.0 meV, $\eta_\text{qc}=0.5$ meV, and $ N_t/t = 2~\text{meV}/\hbar $, in all cases. For all simulations shown in Fig. ~\ref{F3}, the density is initially $n_{0}(0) = 1.0$ and corresponds to the injection of an electron into the chain.  Over time, the densitys at sites $i > 1$ increase, as electron density flows towards the drain as indicated by the dotted green line in Fig.~\ref{F3}a. Black asterisks at $i=1$ correspond to the injection of an electron while white asterisks at $i=L$ correspond to the removal of electrons.  We see that injecting an electron causes a sharp increase in density while the removal of an electron causes a sharp decrease in density, as expected.  When electrons are not immediately removed, they can reflect off of the right end of the wire.  Backpropagation of electron density can be seen as marked by the dotted yellow line.  This results in a peak in density at the left end not corresponding to the injection of an electron.  The flow of density from $i=1$ to $i=7$ is faster for $\gamma$=5.0 meV  (Fig.~\ref{F3}b,d) than for $\gamma$=3.0 meV (Fig.~\ref{F3}a,c), on both the classical and quantum computers. 

While the data taken from the quantum computer looks qualitatively very similar to that taken from the classical computer, there are errors arising from the quantum noise.  In Table~\ref{T2}, we show selected data points from Fig.~\ref{F3}b,d.  We see that the quantum computer tends to smooth the sharp peaks in electron density.  This is likely due to decoherence from the quantum computer.   

\begin{figure}[h!]
\vspace{2mm}
\begin{center}
\includegraphics[width=\columnwidth]{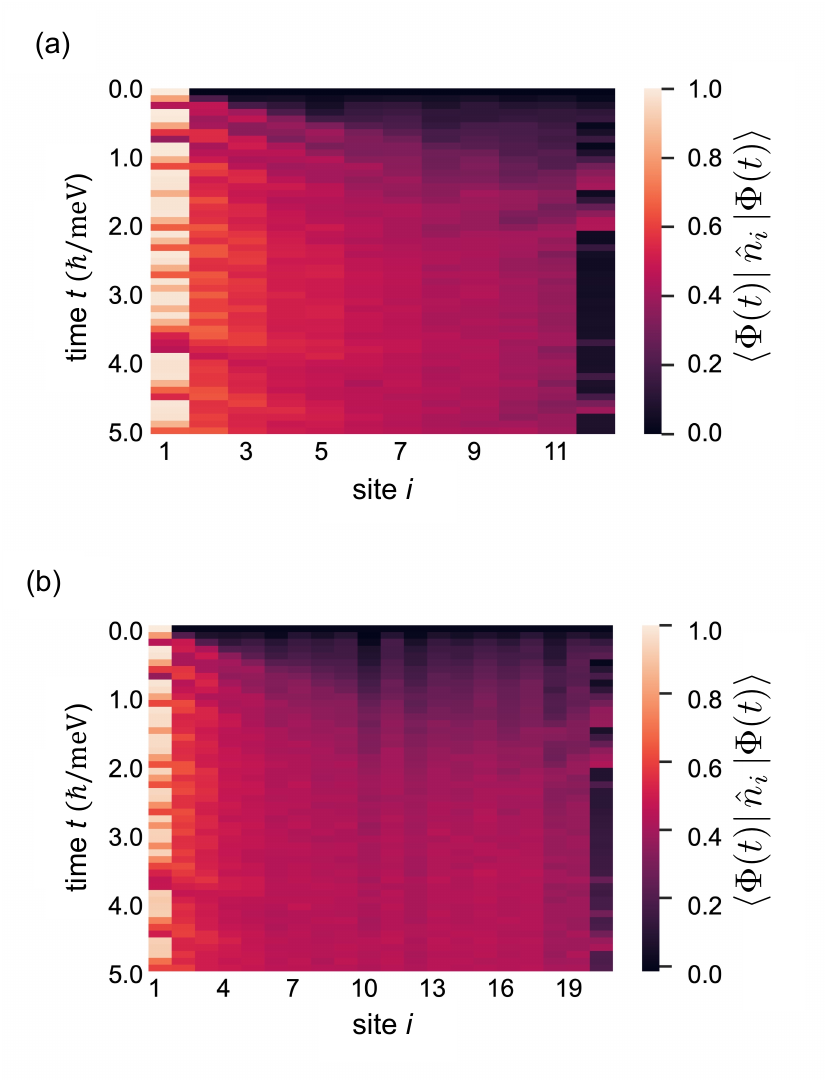}
\end{center}
\vspace{-2mm}
\caption{Simulations on a quantum computer with varying number of sites : (a) L=12, (b) $L=20$. In both cases the source is at $i=1$ and the drain is at $i=L$. }  
\label{F4}
\vspace{-3mm}
\end{figure}

We then increase the number of sites and run these simulations on the quantum computer with $\gamma$=5.0 meV and all other parameters the same as listed above. Figure~\ref{F4} shows the results for these simulation. 
The time evolution of the electron density is qualitatively the same for longer chains with density flowing from the source to the drain.  For longer chains the electron density rises at site $i=L$ at later times, as expected.  While there is a larger degree of noise in Fig.~\ref{F4}b there is still a clear signature of electron movement across the chain, even for this 20-qubit simulation.  

\begin{table}
\begin{tabular}{ | c || c | c | c | c | c | c |  }
\hline
 & \multicolumn{3}{|c|}{Classical} & \multicolumn{3}{|c|}{Quantum} \\
\hline
$t~(\hbar/\text{meV})$ &  $n_1(t)$ & $n_4(t)$ & $n_7(t)$ &  $n_1(t)$ & $n_4(t)$ & $n_7(t)$ \\
\hline
\hline
 0.0 & 1.00 & 0.00 & 0.00 & 1.00 & 0.00 & 0.00 \\ 
 0.5 & 0.74 & 0.27 & 0.00 & 0.68 & 0.34 & 0.00 \\ 
 1.0 & 0.75 & 0.32 & 0.69 & 0.68 & 0.37 & 0.28 \\  
 1.5 & 0.80 & 0.59 & 0.00 & 0.69 & 0.44 & 0.01 \\ 
 2.0 & 0.41 & 0.61 & 0.75 & 0.61 & 0.39 & 0.48 \\   
 2.5 & 1.00 & 0.53 & 0.00 & 0.80 & 0.44 & 0.01 \\ 
 3.0 & 1.00 & 0.52 & 0.00 & 0.84 & 0.44 & 0.01 \\ 
 3.5 & 0.42 & 0.55 & 0.00 & 0.54 & 0.45 & 0.01 \\ 
 4.0 & 1.00 & 0.37 & 0.00 & 0.79 & 0.40 & 0.01 \\ 
 4.5 & 0.33 & 0.49 & 0.23 & 0.49 & 0.42 & 0.12 \\ 
 5.0 & 0.36 & 0.56 & 0.00 & 0.54 & 0.40 & 0.01 \\ 
 \hline
\end{tabular}
\caption{Data from Fig.~\ref{F3}b (Classical) and Fig.~\ref{F3}d (Quantum).}
\label{T2}
\end{table}

\subsection{Conclusions}
We have developed a method for simulating the dynamics of open electronic systems on a quantum computer. This approach entails mid-circuit measurements to inject and drain electrons from the system.  We demonstrate this method on a quantum computer for a model system coupled to a pair of contacts.  Our approach requires fewer qubits than previously described methods and can be used to study the electron dynamics of open systems in general.  
 Future work could consider systems relevant to molecular electronics, i.e., those with a barrier to transfer along the chain.

\section{Acknowledgements} This work has been supported by the Office of Naval Research (ONR) through the U.S. Naval Research Laboratory (NRL). We acknowledge QC resources from IBM through a collaboration with the Air Force Research Laboratory (AFRL).  

\appendix 

\section{Relationship to the Lindblad Equation}

\label{rle}

The goal is to use a quantum computer to reproduce the dynamics of a quantum system weakly coupled an electronic contact.  We know that the dynamics of such a system are well described by the Lindblad master equation
\begin{equation}
    \frac{d\hat \rho}{dt} = -i[\hat H,\hat \rho] +  \sum_{\alpha}  \left( \hat L_{\alpha} \hat \rho \hat L_{\alpha}^{\dagger} - \frac{1}{2} \hat L_{\alpha}^{\dagger} \hat L_{\alpha} \hat \rho - \frac{1}{2} \hat \rho  \hat L_{\alpha}^{\dagger} \hat L_{\alpha}   \right)
\end{equation}
where $\hat H$ is the Hamiltonian of the system, $\hat L_0 = \sqrt{\Gamma f(\mu)}\hat c^{\dagger}_l$ is the Lindblad operator for the incoming electrons, $\hat L_1 = \sqrt{\Gamma [1-f(\mu)]}\hat c_l$ is the Lindblad operator for the outgoing electrons, $\Gamma$ is the tunneling rate between the contact and the system, $f(\mu)$ is the Fermi-Dirac distribution function evaluated at the chemical potential of the lead given that the lead has an electrochemical potential $\mu$, $\hat c^{\dagger}_i$ is the electron creation operator at point $i$, and $l$ is the point of contact.  We use selective mid-circuit measurements to reproduce these dynamics on a quantum computer.

\begin{figure}[h!]
\vspace{2mm}
\begin{center}
\includegraphics[width=\columnwidth]{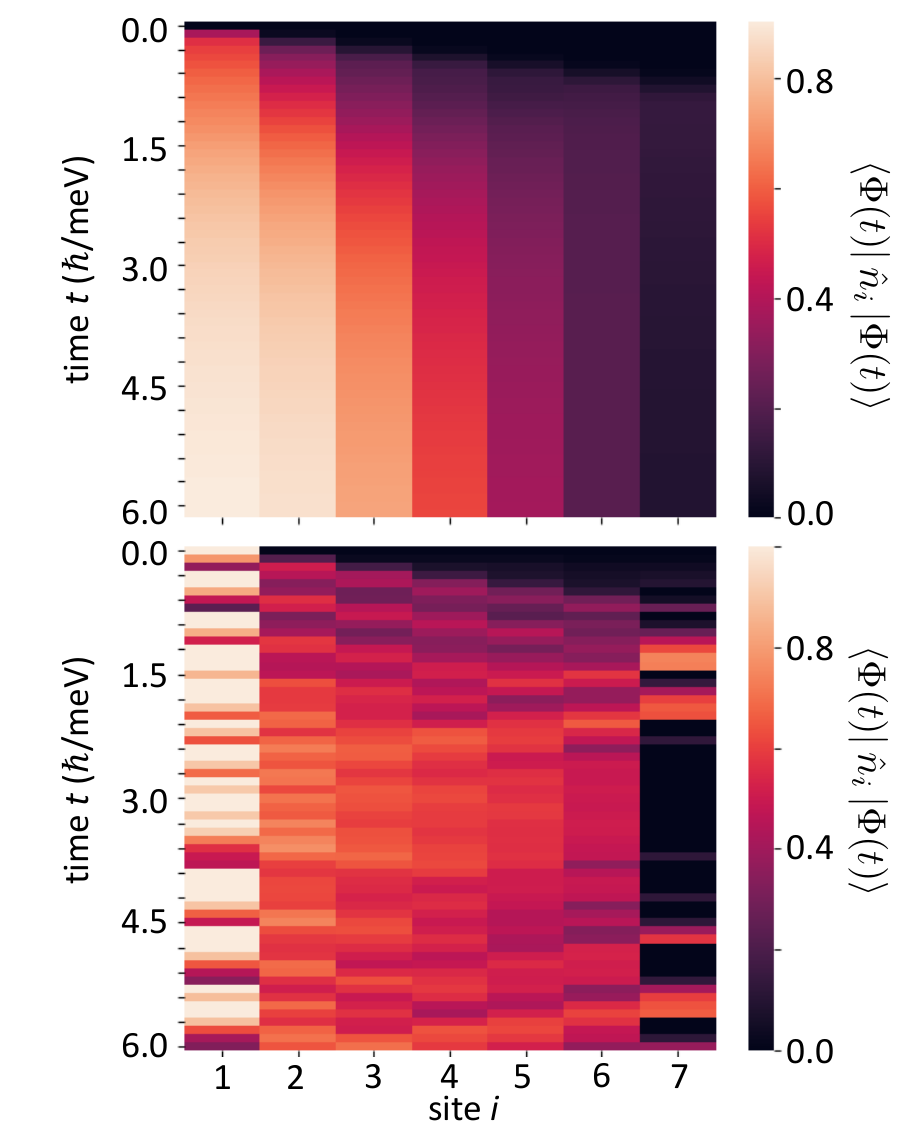}
\end{center}
\vspace{-2mm}
\caption{ Simulations of open-system dynamics using Lindblad evolution (a) and our method (b), for $\gamma$=3.0 meV, and 5.0 meV. The source and drain are at $i=1$ and $i=7$, respectively with $P_{\text{in}} = P_{\text{out}} = 0.5$.  Both approaches use Trotterized time evolution with $N_t/t = 2~\text{meV}/\hbar$}  
\label{FB}
\vspace{-3mm}
\end{figure}

Let us initialize the quantum register to a state $\rho(0)$ and then evolve the system according to
\begin{equation}
    \frac{d \hat \rho(t)}{dt} = -i[\hat H, \hat \rho(t)].
\end{equation} 
After a short amount of time $\delta t$ we randomly perform one of three options: (1) with probability $P_{\text{in}} = r_{\text{in}}\delta t$, we measure point $l$ and if it does not already contain an electron then we create one, (2) with probability $P_{\text{out}} = r_{\text{out}}\delta t$, we measure point $l$ and if it contains an electron we remove it, (3) with probability $(1-P_{\text{in}}-P_{\text{out}})$ we do nothing, where $r_{\text{in}}$ and $r_{\text{out}}$ are rates to be determined later.  After this step, the density matrix is
\begin{equation}
\begin{split}
      \hat \rho^M(t + \delta t) =& (1 - r_{\text{in}}\delta t - r_{\text{out}}\delta t )\hat \rho(t + \delta t)
      \\
      +& r_{\text{in}}\delta t (\hat n_l \hat \rho(t + \delta t) \hat n_l + \hat c^{\dagger}_l\bar n_l \hat \rho(t + \delta t) \bar n \hat c_l ) 
      \\
      +& r_{\text{out}}\delta t (\hat c_l \hat n_l \hat \rho(t + \delta t) \hat n_l \hat c^{\dagger}_l + \bar n_l \hat \rho(t + \delta t) \bar n_l)
\end{split}
\label{pmdt}
\end{equation}
where $\hat n_l = c^{\dagger}_l c_l$ and $\bar n_l = c_l c^{\dagger}_l$.  If we update $\hat \rho$ after the event to $\hat \rho ^M$ and take the limit that $\delta t \rightarrow 0$ we can write the derivative as  
\begin{equation}
\begin{split}
     \frac{d\hat\rho}{dt}\bigg\rvert_{t+\delta t} = \lim_{\delta t\rightarrow 0}\frac{\rho^M(t+dt) - \rho(t)}{\delta t}.
    \label{dpm}
    \\
    \quad
\end{split}
\end{equation}
Inserting Eq.~\eqref{pmdt} into Eq.~\eqref{dpm} we have
\begin{equation}
\begin{split}
     \frac{d\hat\rho}{dt} &= -i[\hat H,\hat\rho] 
     \\
     &+ r_{\text{in}} \left( \hat c^{\dagger}\hat \rho \hat c - \frac{1}{2}\bar n \rho - \frac{1}{2} \hat \rho  \bar n   \right) 
          \\
     &+ r_{\text{out}} \left( \hat c\hat \rho \hat c^{\dagger} - \frac{1}{2} \hat n \hat \rho - \frac{1}{2}   \hat \rho \hat n   \right)
     \\
     &+ r_{\text{in}} \left( \hat n \hat \rho \hat n - \frac{1}{2}\hat n \hat \rho - \frac{1}{2} \rho   \hat n   \right) 
     \\
     &+ r_{\text{out}} \left(  \bar n \hat \rho  \bar n - \frac{1}{2} \bar n \hat \rho - \frac{1}{2} \hat \rho \bar n   \right).    
\end{split}
\label{dpdt}
\end{equation}
where we used $c^{\dagger} \bar n = c^{\dagger}$, $\hat n c^{\dagger}  = c^{\dagger}$, $c \hat n = c$, $ \bar n c = c$, and $\hat \rho = 1/2(\hat n + \bar n )\hat \rho + 1/2 \hat \rho (\hat n + \bar n)$.
This reproduces the Lindblad master equation for $\hat L_0 = \sqrt{r_\text{in}}\hat c^{\dagger}$, $\hat L_1 = \sqrt{r_{\text{out}}} \hat c$, $\hat L_2 = \sqrt{r_\text{in}}\hat n$, and $\hat L_3 = \sqrt{r_{\text{out}}} \bar n$.  For $r_{\text{in}} = \Gamma f(\mu)$ and $r_{\text{out}} = \Gamma [1-f(\mu)]$ the Lindblad operators $\hat L_0$ and $\hat L_1$ represent electron source and drain channels.  The operators $\hat L_3$ and $\hat L_4$ represent depolarizing channels.  These extra depolarising channels are unavoidable.  However, they do not significantly alter the dynamics of the circuit.  Eq.~\eqref{dpdt} can be understood as the dynamics of a quantum system in contact with a noisy quantum lead where the noise in the lead causes some decoherence at the point of contact.

We numerically compare our method to Lindblad evolution in Fig.~\ref{FB} with select data being shown in Table~\ref{TB}.  One striking difference is that, in the Lindblad evolution, the contacts are modeled as continuous injection or removal of fractional electron density.  In our method the contacts probabilistically inject or remove a full electron.  In the limit $\delta t \rightarrow 0$, these methods become indistinguishable, however, even for finite $\delta t$ the two methods share many qualitative features.  Note that the density reaches the right edge at the same time in both methods, indication that the electron group velocity is the same for both methods.  Furthermore, the average electron density in the center of the wire reaches the same value for both methods, at long times.  It should be noted that Trotterization errors may accumulate differently for the two methods, especially because our method requires an extra approximation in order to convert the evolution operator into quantum gates.   

\begin{table}
\begin{tabular}{ | c || c | c | c | c | c | c |  }
\hline
 & \multicolumn{3}{|c|}{Classical} & \multicolumn{3}{|c|}{Lindblad} \\
\hline
$t~(\hbar/\text{meV})$ &  $n_1(t)$ & $n_4(t)$ & $n_7(t)$ &  $n_1(t)$ & $n_4(t)$ & $n_7(t)$ \\
\hline
\hline
 4.0 & 1.00 & 0.51 & 0.00 & 0.87 & 0.50 & 0.09 \\ 
 4.1 & 1.00 & 0.50 & 0.00 & 0.88 & 0.50 & 0.09 \\ 
 4.2 & 1.00 & 0.54 & 0.12 & 0.88 & 0.50 & 0.09 \\ 
 4.3 & 0.90 & 0.55 & 0.00 & 0.88 & 0.51 & 0.09 \\ 
 4.4 & 0.67 & 0.52 & 0.00 & 0.88 & 0.51 & 0.09 \\ 
 4.5 & 0.49 & 0.49 & 0.12 & 0.89 & 0.52 & 0.09 \\ 
 4.6 & 1.00 & 0.51 & 0.38 & 0.89 & 0.52 & 0.09 \\ 
 4.7 & 1.00 & 0.56 & 0.58 & 0.89 & 0.52 & 0.09 \\
 4.8 & 1.00 & 0.52 & 0.00 & 0.89 & 0.53 & 0.09 \\ 
 4.9 & 0.89 & 0.46 & 0.00 & 0.89 & 0.53 & 0.09 \\ 
 5.0 & 0.64 & 0.48 & 0.00 & 0.89 & 0.53 & 0.09 \\ 
 5.1 & 0.45 & 0.44 & 0.00 & 0.89 & 0.53 & 0.09 \\ 
 5.2 & 0.35 & 0.57 & 0.13 & 0.90 & 0.54 & 0.09 \\ 
 5.3 & 1.00 & 0.58 & 0.41 & 0.90 & 0.54 & 0.09 \\ 
 5.4 & 0.88 & 0.55 & 0.59 & 0.90 & 0.54 & 0.08 \\ 
 5.5 & 1.00 & 0.46 & 0.64 & 0.90 & 0.54 & 0.08 \\ 
 5.6 & 1.00 & 0.41 & 0.66 & 0.90 & 0.55 & 0.08 \\ 
 5.7 & 0.89 & 0.53 & 0.00 & 0.90 & 0.55 & 0.08 \\ 
 5.8 & 0.63 & 0.64 & 0.00 & 0.90 & 0.55 & 0.08 \\ 
 5.9 & 0.41 & 0.62 & 0.12 & 0.90 & 0.55 & 0.08 \\ 
 6.0 & 0.33 & 0.59 & 0.38 & 0.90 & 0.55 & 0.08 \\ 
 \hline
 avg. & 0.78 & 0.53 & 0.20 & 0.89 & 0.53 & 0.09 \\
 \hline
\end{tabular}
\caption{Data from Fig.~\ref{FB}a (Lindblad) and Fig.~\ref{FB}b (Classical).}
\label{TB}
\end{table}

\section{Device Specifications}
\label{apexb}

The quantum computing is performed on IBM's Heron Processor named $imb\_torino$.  This is a 133 qubit device with a median T1 time of $186.63~\mu s$ and a median T2 time of $139.74~\mu s$.  The basis gates for $imb\_torino$ are the controlled-Z gate (CZ) and the single qubit rotations (sx) and (rz).  The CZ gates have a median error of $3.8\times10^{-3}$ and the sx gates have a median error of $2.9\times10^{-4}$.  The rz gates are implemented in software as phase shifts~\cite{McKay2017} and have no associated error.  The median readout assignment error is $2.3\times10^{-2}$ and the median gate time is $68~ns$.  Full details of the device can be found on the IBM Quantum Platform~\cite{IBM_web}.

\pagebreak

\bibliography{ref}


\end{document}